%% file: draft.tex
\documentclass[runningheads]{llncs}
\usepackage{graphicx}
\usepackage{amsmath, amssymb, array, booktabs, multicol, url}

\usepackage[inline]{enumitem}

\usepackage[vlined, ruled, lines numbered]{algorithm2e}

\newcommand\blfootnote[1]{%
  \begingroup
  \renewcommand\thefootnote{}\footnote{#1}%
  \addtocounter{footnote}{-1}%
  \endgroup
}

\newcolumntype{L}[1]{>{\raggedright \let\newline\\\arraybackslash\hspace{0pt}}m{#1}}
\newcolumntype{C}[1]{>{\centering   \let\newline\\\arraybackslash\hspace{0pt}}m{#1}}
\newcolumntype{R}[1]{>{\raggedleft  \let\newline\\\arraybackslash\hspace{0pt}}m{#1}}

\input{shortcuts.tex}

\begin{document}
\title{Parallel Robust Computation of Generalized Eigenvectors of Matrix Pencils}
\titlerunning{Generalized Eigenvectors of Matrix Pencils}
%
\author{Carl Christian Kjelgaard Mikkelsen\orcidID{0000-0002-9158-1941} \\ Mirko Myllykoski\orcidID{0000-0002-3689-0899}}
\authorrunning{C.C. Kjelgaard Mikkelsen \and M. Myllykoski}
%
\institute{Department of Computing Science and HPC2N \\ Ume{\aa} University \\ 90187 Ume{\aa}, Sweden \\ \email{\{spock,mirkom\}@cs.umu.se}}
\maketitle              

\begin{abstract} In this paper we consider the problem of computing generalized eigenvectors of a matrix pencil in real Schur form. In exact arithmetic, this problem can be solved using substitution. In practice, substitution is vulnerable to floating-point overflow. The robust solvers \texttt{xTGEVC} in LAPACK prevent overflow by dynamically scaling the eigenvectors. These subroutines are sequential scalar codes which compute the eigenvectors one by one. In this paper we discuss how to derive robust blocked algorithms. The new StarNEig library contains a robust task-parallel solver \texttt{Zazamoukh} which runs on top of StarPU. Our numerical experiments show that \texttt{Zazamoukh} achieves a super-linear speedup compared with \texttt{DTGEVC} for sufficiently large matrices.
\end{abstract}

\blfootnote{This manuscript was accepted to \emph{13th International Conference on Parallel Processing and Applied Mathematics} (PPAM2019), Bialystok, Poland, September 8--11, 2019.
The final authenticated version is available online at \\\url{https://doi.org/10.1007/978-3-030-43229-4_6}.}

\section{Introduction} \label{sec:introduction}

Let $A \in \Rmm$ and let $B \in \Rmm$. The matrix pencil $(A,B)$ consists of all matrices of the form $A - \lambda B$ where $\lambda \in \C$. The set of (generalized) eigenvalues of the matrix pencil $(A,B)$ is given by
\be
\lambda(A,B) = \{ \lambda \in C \: : \det(A-\lambda B) = 0 \}.
\ee
We say that $z \in \Cm$ is a (generalized) eigenvector of the matrix pencil $(A,B)$ if and only if
$z \not = 0$ and
\be
Az = \lambda Bz.
\ee
The eigenvalues of $(A,B)$ can be computed by first reducing $(A,B)$ to real Schur form $(S,T)$. Specifically, there exist orthogonal matrices $U$ and $V$ such that $S = U^TAV$ is quasi-upper triangular and $T = U^T B V$ is upper triangular. It is clear that
\be
\lambda(A,B) = \lambda(S,T).
\ee
Moreover, if $z$ is a generalized eigenvector of $(S,T)$ corresponding to the eigenvalue $\lambda$, then $Uz$ is a generalized eigenvector of $(A,B)$ corresponding to the eigenvalue $\lambda$.

In this paper, we consider the parallel computation of eigenvectors of a matrix pencil in real Schur form.
In exact arithmetic, this problem can be solved using substitution. However, substitution is very vulnerable to floating-point overflow.

In LAPACK \cite{laug} there exists a family \texttt{xTGEVC} of subroutines which compute the generalized eigenvectors of a matrix pencil in Schur form.
They prevent overflow by dynamically scaling the eigenvectors.
These subroutines are scalar codes which compute the eigenvectors one by one.
In this paper we discuss the construction of algorithms which are not only robust, but blocked and parallel.

Our paper is organized as follows. In Section \ref{sec:real} we consider the problem of computing the eigenvectors of a matrix pencil in real Schur form using real arithmetic. This problem is equivalent to solving a homogeneous matrix equation of the form
\be \label{equ:central-matrix-equation}
SV\!D - TV\!B = 0
\ee
where $D$ is diagonal and $B$ is block diagonal with diagonal blocks which are either 1-by-1 or 2-by-2.
In Section \ref{sec:block} we present a blocked algorithm for solving this matrix equation.
In Section \ref{sec:robust} we discuss how to prevent overflow in this algorithm.
The concept of an augmented matrix is central to this discussion.
A robust task-parallel solver {\tt Zazamoukh} has been developed and integrated into the new StarNEig library for solving non-symmetric eigenvalue problems \cite{starneig1,starneig2}.
The performance of {\tt Zazamoukh} is compared to LAPACK in Section \ref{sec:experiments}.
We outline several directions for future work in Section \ref{sec:conclusion}.

\section{Real arithmetic} \label{sec:real}

Let $A \in \Rmm$ and $B \in \Rmm$ be given. The set of generalized eigenvalues of the matrix pencil $(A,B)$ can be computed by first reducing $(A,B)$ to generalized real Schur form. Specifically, there exist orthogonal matrices $U$ and $V$ such that
\bes
S = U^T A V =
\bbm S_{11} & S_{12} & \dots & S_{1p} \\ & S_{22} & \dots & S_{2p} \\ & & \ddots & \vdots \\ & & & S_{pp} \ebm, \quad T = U^T B V = \bbm T_{11} & T_{12} & \dots & T_{1p} \\ & T_{22} & \dots & T_{2p} \\ & & \ddots & \vdots \\ & & & T_{pp} \ebm
\ees
are upper block triangular and $\dim(S_{jj}) = \dim(T_{jj}) \in \{1,2\}$.
It is clear that
\bes
\lambda(S,T) = \cup_{j=1}^p \lambda(S_{jj}, T_{jj}).
\ees
In order to simplify the discussion, we will make the following assumptions.
\begin{enumerate}
\item If $\dim(S_{jj}) = 1$, then $(S_{jj}, T_{jj})$ has a single real eigenvalue.
\item If $\dim(S_{jj}) = 2$, then $(S_{jj}, T_{jj})$ has two complex conjugate eigenvalues.
\item All eigenvalues are distinct.
\end{enumerate}
We follow the standard convention and represent eigenvalues $\lambda$ using an ordered pair $(\alpha, \beta)$ where $\alpha \in \C$ and $\beta \ge 0$. If $\beta > 0$, then $\lambda = \alpha/\beta$ is a finite eigenvalue. The case of $\alpha \in \R - \{0\}$ and $\beta = 0$, corresponds to an infinite eigenvalue. The case of $\alpha = \beta = 0$ corresponds to an indefinite problem.

We now consider the problem of computing generalized eigenvectors of $(S,T)$. Our goal is to obtain an equivalent problem which can be solved using a blocked algorithm.

\subsection{Computing a single eigenvector}

It this subsection, we note that the problem of computing a single generalized eigenvector of $(S,T)$ is equivalent to solving a tall homogenous matrix equation involving real matrices.
Let $\lambda \in \lambda(S_{jj},T_{jj})$ and let $\lambda = \frac{\alpha_j}{\beta_j}$ where $\beta_j > 0$ and
\bes
\alpha_j = a_j + i b_j \in \C.
\ees
Let $m_j = \dim(S_{jj})$ and let $D_{jj} \in \R^{m_j \times m_j}$ and $B_{jj} \in \R^{m_j \times m_j}$ be given by
\begin{equation} \label{equ:DB1}
  D_{jj} = \beta_j, \quad B_{jj} = a_j
\end{equation}
when $\dim(S_{jj}) = 1$ (or equivalently $b_j = 0$) and
\begin{equation} \label{equ:DB2}
  D_{jj} =
  \begin{bmatrix}
    \beta_j & 0 \\ 0 & \beta_j
  \end{bmatrix},
  \quad
  B_{jj} =
  \begin{bmatrix}
    a_j & b_j \\
    -b_j & a_j
  \end{bmatrix}
\end{equation}
when $\dim(S_{jj}) = 2$ (or equivalently $b_j \not = 0$). With this notation, the problem of computing an eigenvector is equivalent to solving the homogeneous linear equation
\begin{equation} \label{equ:tall-matrix-equation}
  SV\!D_{jj} - TV\!B_{jj} = 0
\end{equation}
with respect to $V \in \R^{m \times n_j}$. This is an immediate consequence of the following lemma.

\begin{lemma} Let $\lambda \in \lambda(S_{jj}, T_{jj})$ and let $\lambda = (a_j + ib_j)/\beta$ where $\beta > 0$. Then the following statements are true.
  \begin{enumerate}
  \item If $\dim(S_{jj}) = 1$, then $x \in \Rm$ is a real eigenvector corresponding to the real eigenvalue $\lambda \in \R$ if and only if $V = \bbm x \ebm$ has rank 1 and solves equation \eqref{equ:tall-matrix-equation}.
  \item If $\dim(S_{jj}) = 2$, then $z = x + iy \in \Cm$ is a complex eigenvector corresponding to the complex eigenvalue $\lambda \in \C$ if and only if $V = \bbm x & y \ebm$ has rank 2 and solves equation \eqref{equ:tall-matrix-equation}.
  \end{enumerate}
\end{lemma}

\subsection{Computing all eigenvectors}
It this subsection, we note that the problem of computing all generalized eigenvectors of $(S,T)$ is equivalent to solving a homogenous matrix equation involving real matrices.
Specifically, let $D \in \Rmm$ and $B \in \Rmm$ be given by
\be
D = \diag\{D_{11}, D_{22}, \dotsc, D_{pp}\}, \quad B = \diag\{B_{11}, B_{22}, \dotsc, B_{pp}\}.
\ee
where $D_{jj}$ and $B_{jj}$ are given by equations \eqref{equ:DB1} and \eqref{equ:DB2}.
Then $V = \bbm V_1 & V_2 & \dots V_p \ebm$ solves the homogeneous matrix equation
\begin{equation} \label{equ:large-matrix-equation}
  SV\!D - TV\!B = 0
\end{equation}
if and only if $V_j \in \R^{m \times n_j}$ solves equation \eqref{equ:tall-matrix-equation}.

\section{A blocked algorithm} \label{sec:block}

It is straightforward to derived a blocked algorithm for solving the homogeneous matrix equation \eqref{equ:large-matrix-equation}. Specifically, \emph{redefine}
\bes
S = \bbm S_{ij} \ebm, \quad i, j \in \{1,2,\dots,M\}
\ees
to denote \textit{any} partitioning of $S$ into an $M$ by $M$ block matrix which does not split any of the 2-by-2 blocks along the diagonal of $S$. Apply the same partitioning to $T$, $D$, $B$, and $V$. Then $S$, $T$, and $V$ are block triangular. It is straightforward to see that equation \eqref{equ:large-matrix-equation} can be solved using Algorithm \ref{alg:blocked-computation-all-eigenvectors}.

\begin{algorithm}[htbp]
  \caption{Blocked computation of all generalized eigenvectors} \label{alg:blocked-computation-all-eigenvectors}
  \For{$j \gets 1,2,\dotsc,p$}{
    Compute generalized eigenvectors $Y_{jj}$ for $(S_{jj},T_{jj})$\;
    \For{$i \gets j-1,\dotsc,1$}{
      \For{$k \gets 1,2,\dotsc, i$}{
        $Y_{kj} = Y_{kj} - (S_{k,i+1} Y_{i+1,j} D_{jj} - T_{k,i+1} Y_{i+1,j} B_{jj})$ 
      }
      Solve
      \be
      S_{ii} Z D_{jj} - T_{ii} Z B_{jj} = Y_{ij}
      \ee
      with respect to $Z$ and set $Y_{ij} \gets Z$\;
    }
  }
\end{algorithm}
We see that Algorithm \ref{alg:blocked-computation-all-eigenvectors} can be implemented using three distinct kernels:
\begin{enumerate}
\item \label{kernel:eig} Compute the eigenvectors corresponding to a small matrix pencil $(S,T)$ in real Schur form.
\item \label{kernel:update} Execute linear updates of the form
  \be \label{equ:update}
  Y \gets Y - (S X D - T X B)
  \ee
  where $S$ and $T$ are small dense matrices, $D$ is diagonal and $B$ is block diagonal with blocks of size $1$ or $2$.
\item \label{kernel:solve} Solution of small matrix equations of the form
  \be \label{equ:solve}
  S Z D - T Z B = Y
  \ee
  where $(S,T)$ is a matrix pencil in real Schur form and $D$ is diagonal and $B$ is block diagonal with diagonal blocks of dimension $1$ or $2$.
\end{enumerate}
Once these kernels have been implemented, it is straightforward to parallelize Algorithm \ref{alg:blocked-computation-all-eigenvectors} using a task-based runtime system such as StarPU \cite{StarPU}. 

\section{Deriving a robust blocked algorithm} \label{sec:robust}

The subroutines {\tt xTGEVC} prevent overflow using principles originally derived by Edward Anderson and implemented in the subroutines {\tt xLATRS} \cite{lawn36}. These subroutines apply to triangular linear systems
\be \label{equ:triangular-system}
Tx=b
\ee
with a single right-hand side $b$. Mikkelsen and Karlsson \cite{ppam2017} formalized the work of Anderson and derived a robust blocked algorithm for solving \eqref{equ:triangular-system}. Mikkelsen, Schwarz and Karlsson \cite{ccpe2018} derived a robust blocked algorithm for solving triangular linear systems $$TX=B$$ with multiple right hand sides. Their StarPU implementation ({\tt Kiya}) is significantly faster than {\tt DLATRS} when numerical scaling is necessary and not significantly slower than {\tt DTRSM} when numerical scaling is unnecessary.

A robust variant of Algorithm \ref{alg:blocked-computation-all-eigenvectors} can be derived using
the principles applied by Mikkelsen, Schwarz and Karlsson \cite{ccpe2018}. We can use the two real functions {\tt ProtectUpdate} and {\tt ProtectDivision} introduced by Mikkelsen and Karlsson \cite{ppam2017}. They are used to prevent scalar divisions and linear updates from exceeding the overflow threshold $\Omega > 0$. They have the following key properties.

\begin{enumerate}
\item If $t \not = 0$ and $|b| \leq \Omega$, and
  \be
  \xi = {\tt ProtectDivision}(|b|,|t|)
  \ee
  then $\xi \in (0,1]$ and $|\xi b| \leq |t| \Omega$. It follows that the scaled division
  \be
  y \gets \frac{(\xi b)}{t}
  \ee
  cannot exceed $\Omega$.
\item If $Z = Y - TX$ is defined, with
  \be
  \|T\|_\infty \leq \Omega, \quad \|X\|_\infty \leq \Omega, \quad \|Y\|_\infty \leq \Omega,
  \ee
  and
  \be
  \xi = {\tt ProtectUpdate}(\|Y\|_\infty, \|T\|_\infty, \|X\|_\infty)
  \ee
  then $\xi \in (0,1]$ and $\xi (\|Y\|_\infty + \|T\|_\infty \|X\|_\infty) \leq \Omega$. It follows that
  \be
  Z \gets (\xi Y) - T (\xi X) = (\xi Y) - (\xi T) X
  \ee
  can be computed without any intermediate or final result exceeding $\Omega$.
\end{enumerate}

Now consider the kernels needed for implementing Algorithm \ref{alg:blocked-computation-all-eigenvectors}. It is easy to understand that they can be implemented using loops, divisions and linear updates. It is therefore plausible that robust variants can be implemented using {\tt ProtectDivision} and {\tt ProtectUpdate}. However, resolving all the relevant details requires many pages. Here we explain how the updates given by equation \eqref{equ:update} can be done without exceeding the overflow threshold $\Omega$. We will use the concept of an \emph{augmented} matrix introduced by Mikkelsen, Schwarz and Karlsson \cite{ccpe2018}.

\begin{definition} Let $X \in \Rmn$ be partitioned into $k$ block columns
  \be
  X =  \bbm X_1 & X_2 & \dotsc & X_k \ebm, \quad X_j \in \R^{m \times n_j}, \quad \sum_{j=1}^k n_j = n,
  \ee
  and let $\alpha \in \R^k$ have $\alpha_j \in (0,1]$. The augmented matrix $\langle \alpha, X \rangle$ represents the real matrix $Y$ given by
  \be
  Y = \bbm Y_1 & Y_2 & \dotsc & Y_k \ebm, \quad Y_j = \alpha_j^{-1} Y_j.
  \ee
\end{definition}

This is a trivial extension of the original definition which only considered the case of $k=n$. The purpose of the scaling factors $\alpha_j$ is used to extend the normal representational range of our floating-point numbers. 

\begin{algorithm} \caption{Right updates with block diagonal matrix} \label{alg:right}
  \KwData{An augmented matrix $\av{\alpha}{X}$ where
    \bes
    X =  \bbm X_1 & X_2 & \dotsc & X_k \ebm, \quad \|X_j\|_\infty \leq \Omega, \quad \alpha \in \R^k
    \ees
    and matrices $B_j$ such that $\|B_j\|_\infty \leq \Omega$ and $Y_j = X_j B_j$ is defined.}
  \KwResult{An augmented matrix $\av{\beta}{Y}$ such that
    \be
    \beta_j^{-1} Y_j = (\alpha_j^{-1} X_j) B_j, \quad \|Y_j \|_\infty \leq \Omega,
    \ee
    and $Y$ can be computed without exceeding $\Omega$.
  }
  \For{$j=1,2\dots,k$}{
    $\gamma_j = \PU(\|X_j\|_\infty, \|B_j\|_\infty, 0)$\;
    $Y_j = (\gamma_j X_j) B_j$\;
    $\beta_j = \alpha_j \gamma_j$
  }
\end{algorithm}

Now $Z_1 = XD$ can be computed without overflow using Algorithm \ref{alg:right}. In our case $D$ is diagonal, so we are merely scaling the columns of $X$. However, $Z_2 = Y - SZ_1$ can now be computed without overflow using Algorithm \ref{alg:left}. It follows that $Y \gets (Y - S(XD)) + T(XB)$ can be computed without overflow using two applications of Algorithm \ref{alg:right} and Algorithm \ref{alg:left}. Now suppose that $Y$ is $m$ by $n$. Then Algorithm \ref{alg:right} does $O(mn)$ flops on $O(mn)$ data. However, Algorithm \ref{alg:left} then does $O(mn^2)$ flops on the \emph{same} data. Therefore, the overall arithmetic intensity is $O(n)$. 

\begin{algorithm} \caption{Left update with dense matrix} \label{alg:left}
  \KwData{A matrix $T$ and augmented matrices $\av{\alpha}{X}$ and $\av{\beta}{Y}$ where
    \be
    X =  \bbm X_1 & X_2 & \dotsc & X_k \ebm, \quad Y = \bbm Y_1 & Y_2 & \dotsc & Y_k \ebm,
    \ee
    such that $Z_j = Y_j - TX_j$ is defined and
    \be
    \|X_j\|_\infty \leq \Omega, \quad  \|Y_j\|_\infty \leq \Omega.
    \ee
  }
  \KwResult{An augmented matrix $\av{\zeta}{Z}$ such that
      \begin{equation}
        \zeta_j^{-1} Z_j = \beta_j^{-1} Y_j - T (\alpha_j^{-1} X_j), \quad \|Z_j\|_\infty \leq \Omega,
      \end{equation}
      and $Z$ can be computed without exceeding $\Omega$.
    }
    \For{$j=1,\dotsc,k$}{
      $\gamma_j = \min \{\alpha_j, \beta_j\}$\;
      $\delta_j = \PU(\|T\|_\infty, (\gamma_j/\alpha_j) \|X_j\|_\infty, (\gamma_j/\beta_j)\|Y_j\|_\infty)$\;
      $X_j \gets \delta_j (\gamma_j/\alpha_j) X_j$\;
      $Y_j \gets \delta_j (\gamma_j/\beta_j) Y_j$\;
      $\zeta_j = \epsilon_j \delta_j \gamma_j$\;
    }
    $Z \gets Y - TX$\;
    \Return $\av{\zeta}{Z}$
\end{algorithm}

In order to execute all linear updates needed for a robust variant of Algorithm \ref{alg:blocked-computation-all-eigenvectors} we require certain norms. Specifically, we need the infinity norms of all super-diagonal blocks of $S$ and $T$. Moreover, we require the infinity norm of certain submatrices of $Y$. These submatrices consists of either a single column (segment of real eigenvector) or two adjacent columns (segment of complex eigenvector). The infinity norm must be computed whenever a submatrix has been initialized or updated. {\tt ProtectUpdate} requires that the input arguments are bounded by $\Omega$ and failure is possible if they are not. It is necessary to scale the matrices $S$ and $T$ to ensure that all blocks have infinity norm bounded by $\Omega$.

\section{Zazamoukh - a task-parallel robust solver} \label{sec:Zazamoukh}

The new StarNEig library runs on top of StarPU and can be used to solve dense non-symmetric eigenvalue problems. A robust variant of Algorithm \ref{alg:blocked-computation-all-eigenvectors} has been implemented in StarNEig. This implementation ({\tt Zazamoukh}) uses augmented matrices and scaling factors which are signed 32 bit integers. \texttt{Zazamoukh} can compute eigenvectors corresponding to a subset $E \subseteq \lambda(S,T)$ which is closed under complex conjugation. 
\texttt{Zazamoukh} is currently limited to shared memory, but an extension to distributed memory is planned.

\subsection{Memory layout}

Given block sizes $mb$ and $nb$ {\tt Zazamoukh} partitions $S$, $T$ and the matrix of eigenvectors $Y$ conformally by rows and columns. In the absence of any 2-by-2 diagonal blocks on the diagonal blocks the tiles of $S$ and $T$ are $mb$ by $mb$ and the tiles of $Y$ are $mb$ by $nb$. The only exceptions can be found along the right and lower boundaries of the matrices. This default configuration is adjusted minimally to prevent splitting any 2-by-2 block of $S$ or separating the real part and the imaginary part of a complex eigenvector into separate block columns.

\subsection{Tasks}

{\tt Zazamoukh} relies on four types of tasks

\begin{enumerate}
\item Pre-processing tasks which compute all quantities needed for robustness. This includes the infinity norm of all super-diagonal tiles of $S$ and $T$ as well as all norms needed for the robust solution of equations of the type \eqref{equ:solve}. If necessary, the matrics $S$ and $T$ are scaled minimally.
  
\item Solve tasks which use {\tt DTGEVC} to compute the lower \emph{tips} of eigenvectors and a robust solver based on {\tt DLALN2} to solve equations of the type \eqref{equ:solve}.
\item Update tasks which execute updates of the type \eqref{equ:update} robustly.
\item Post-processing tasks which enforce a consistent scaling on all eigenvectors.
\end{enumerate}

\subsection{Task insertion order and priorities}

\texttt{Zazamoukh} is closely related to \texttt{Kiya} which solves triangular linear systems with multiple right-hand sides. Apart from the pre-processing and post-processing tasks, the main task graph is a the disjoint union of $p$ task-graphs, one for each block column of the matrix of eigenvectors. {\tt Zazamoukh} uses the same task insertion order and priorities as {\tt Kiya} to process each of the $p$ sub-graphs.

\section{Numerical experiments} \label{sec:experiments}
In this section we give the result of a set of experiments involving tiny ($m\leq 10\,000$) and small ($m\leq 40\,000$) matrices. Each experiment consisted of computing all eigenvectors of the matrix pencil. The run time was measured for (DTGEVC) LAPACK and {\tt Zazamoukh}. Results related to somewhat larger matrices $(m\leq 80\,000)$ can be found in the NLAFET Deliverable 2.7 \cite{D27}.

The experiments were executed on an Intel Xeon E5-2690v4 (“Broadwell”) node with 28 cores arranged in two NUMA islands with 14 cores each. The theoretical peak performance in double-precision arithmetic
is 41.6 GFLOPS/s for one core and 1164.8 GFLOPS/s for a full node.

We used the StarNEig test-program {\tt starneig-test} to generate reproducible experiments. The default parameters produce matrix pencils where approximately 1 percent of all eigenvalues are zeros, 1 percent of all eigenvalues are infinities and there are no indefinite eigenvalues. {\tt Zazamoukh} used the default tile size $mb$ which is 1.6 percent of the matrix dimension for matrix pencils with dimension $m \ge 1000$.

All experiments were executed with exclusive access to a complete node (28 cores). LAPACK was run in sequential mode, while {\tt Zazamoukh} used 28 StarPU workers and 1 master thread. The summary of our results are given in Figure \ref{fig:results}. The speedup of {\tt Zazamoukh} over LAPACK is initially very modest as there is not enough tasks to keep 28 workers busy, but it picks up rapidly and {\tt Zazamoukh} achieves a superlinear speedup over {\tt DTGEVC} when $m \ge 10\,000$. This is an expression of the fact that {\tt Zazamoukh} uses a blocked algorithm, whereas {\tt DTGEVC} computes the eigenvectors one by one. 

\begin{table}
  \centering 
  \begin{tabular}{R{1.5cm}C{0.3cm}R{0.9cm}R{0.9cm}R{0.9cm}C{0.3cm}R{1.5cm}R{1.5cm}C{0.3cm}R{1.5cm}}

    dimension && \multicolumn{3}{c}{eigenvalue analysis} && \multicolumn{2}{c}{runtime (ms)} && SpeedUp \\
    \cmidrule{1-1} \cmidrule{3-5} \cmidrule{7-8} \cmidrule{10-10}
    m && zeros & inf. & indef. && LAPACK & StarNEig && \\
    1000  &&  11 &   13 &  0   &&       295  &     175  &&      1.6857 \\
    2000  &&  25 &   16 &  0   &&      1598  &     409  &&      3.9071 \\
    3000  &&  24 &   30 &  0   &&      6182  &     929  &&      6.6545 \\
    4000  &&  42 &   49 &  0   &&     15476  &    1796  &&      8.6169 \\
    5000  &&  54 &   37 &  0   &&     30730  &    2113  &&     14.5433 \\
    6000  &&  61 &   64 &  0   &&     53700  &    2637  &&     20.3641 \\
    7000  &&  67 &   64 &  0   &&     84330  &    3541  &&     23.8153 \\
    8000  &&  56 &   69 &  0   &&    122527  &    4769  &&     25.6924 \\
    9000  &&  91 &   91 &  0   &&    171800  &    6189  &&     27.7589 \\
    10000 && 108 &   94 &  0   &&    242466  &    7821  &&     31.0019 \\
    20000 && 175 &  197 &  0   &&   2034664  &   49823  &&     40.8378 \\
    30000 && 306 &  306 &  0   &&   7183746  &  162747  &&     44.1406 \\
    40000 && 366 &  382 &  0   &&  17713267  &  380856  &&     46.5091 \\
    \bottomrule
  \end{tabular} \caption{Comparison of sequential DTGEVC to task-parallel {\tt Zazamoukh} using 28 cores. The runtimes are given in milli-seconds (ms). The last column gives the speedup of {\tt Zazamoukh} over LAPACK. Values above 28 correspond to super-linear speedup. All eigenvectors were computed with a relative residual less than $2u$, where $u$ denotes the double precision unit roundoff.} \label{fig:results}
\end{table}

\section{Conclusion} \label{sec:conclusion}

Previous work by Mikkelsen, Schwarz and Karlsson has shown that triangular linear systems can be solved in parallel without overflow using augmented matrices. In this paper we have shown that the eigenvectors of a matrix pencil can be computed in parallel without overflow using augmented matrices. Certainly, robust algorithms are slower than non-robust algorithms when numerical scaling is not needed, but robust algorithms will always return a result which can be evaluated in the context of the user's application. To the best of our knowlegde StarNEig is the only library which contains a parallel robust solver for computing the generalized eigenvectors of a dense nonsymmetric matrix pencil. The StarNEig solver ({\tt Zazamoukh}) runs on top of StarPU and uses augmented matrices and scaling factors with are integer powers of $2$ to prevent overflow. It acheives superlinear speedup compared with (DTGEVC) from LAPACK.
In the immediate future we expect to pursue the following work:
\begin{enumerate}
\item Extend {\tt Zazamoukh} to also compute left eigenvectors. Here the layout of the loops is different and we must use the 1-norm instead of the infinity norm when executing the overflow protection logic.
\item Extend {\tt Zazamoukh} to distributed memory machines.
\item Extend {\tt Zazamoukh}'s solver to use recursive blocking to reduce the run-time further. The solve tasks all lie on the critical path of the task graph. 
\item Extend {\tt Zazamoukh} to complex data-types. This case is simpler than real arithmetic because there are no $2$-by-$2$ blocks on the main diagonal of $S$.
\item Revisit the complex division routine {\tt xLADIV} \cite{baudin2012} which is the foundation for the {\tt DLALN2} routine used by {\tt Zazamoukh}'s solve tasks. In particular, the failure modes of {\tt xLADIV} have not be characterized \cite{Baudin}.
\end{enumerate}


\subsection*{Acknowlegdements}

This work is part of a project (NLAFET) that has received funding from the European Union’s Horizon 2020 research and innovation programme under grant agreement No 671633. This work was supported by the Swedish strategic research programme eSSENCE. We thank the High Performance Computing Center North (HPC2N) at Ume{\aa} University for providing computational resources and valuable support during test and performance runs.

\end{document}

%% file: shortcuts.tex
\newcommand{\R}{\mathbb{R}}

\newcommand{\Rm}{\R^m}
\newcommand{\Rmm}{\R^{m \times m}}
\newcommand{\Rmn}{\R^{m \times n}}

\newcommand{\C}{\mathbb{C}}

\newcommand{\Cm}{\C^m}

\newcommand{\av}[2]{\langle #1, #2 \rangle}

\newcommand{\bes}{\begin{equation*}}
\newcommand{\ees}{\end{equation*}}
\newcommand{\be}{\begin{equation}}
\newcommand{\ee}{\end{equation}}

\newcommand{\diag}{\text{diag}}

\newcommand{\bbm}{\begin{bmatrix}}
  \newcommand{\ebm}{\end{bmatrix}}

\newcommand{\PU}{\texttt{ProtectUpdate}}